\documentclass{PoS}

\usepackage{epsfig}
\usepackage{dcolumn}% Align table columns on decimal point
\usepackage{bm}% bold math
\usepackage{amsmath}
\usepackage{graphicx,comment}

\newcommand{\beq}{\begin{equation}}
\newcommand{\eeq}{\end{equation}}

\title{Pure gauge glueballs at finite temperature }

\ShortTitle{Pure gauge glueballs at finite temperature }
\author{CLQCD Collaboration: \speaker{X.-F. Meng}${}^{ab}$, G. Li${}^{cd}$, Y. Chen${}^{c}$, C. Liu${}^{e}$, Y.-B. Liu${}^a$, J.-P. Ma${}^f$,
 and J.-B. Zhang${}^g$   \\
${}^a$School of Physics, Nankai University, Tianjin 300071, China\\
${}^b$National Supercomputing Center, Tianjin 300457, China\\
${}^c$Institute of High Energy Physics, Chinese Academy of Sciences, Beijing 100049, China \\
${}^d$Department of Physics, Qufu Normal University, Shandong 273165, China \\
${}^e$School of Physics, Peking University, Beijing 100871, China \\
${}^f$Institute of Theoretical Physics, Chinese Academy of Sciences, Beijing 100080, China \\
${}^g$Department of Physics, Zhejiang University, Hangzhou 310027, China\\
E-mail: \email{mengxf@mail.nankai.edu.cn}}

\abstract{ Pure gauge glueballs at finite temperature are investigated in a large temperature range from $0.3T_c$ to
$1.9T_c$ on anisotropic lattices. Optimized glueball operators are used to obtain better signals. It is found in all 20
symmetry channels that the pole masses $M_G$ of glueballs remain almost constants when the temperature approaches the
critical temperature $T_c$ from below, and start to reduce gradually with the temperature going above $T_c$. The
glueball correlators in $0^{++}$, $0^{-+}$, and $2^{++}$ channels, are also analyzed based on the Breit-Wigner ansatz
by assuming a thermal width $\Gamma$ to the pole mass $\omega_0$. While $\omega_0$'s are insensitive to $T$ in the
whole temperature range, $\Gamma$'s exhibit distinct behavior below and above $T_c$: They are only few percents of
$\omega_0$ when $T<T_c$, but grow abruptly when $T>T_c$ and reach values of roughly $\Gamma\sim \omega_0/2$ at
$T\approx 1.9T_c$. }

\FullConference{The XXVII International Symposium on Lattice Field Theory \\
July 26 - 31, 2009\\
Peking University, Beijing, China}

\begin{document}

\section{Introduction}
Quantum Chromodynamics(QCD) is believed to be the fundamental theory of strong interaction. At finite temperature, QCD
is usually described by two extreme pictures. One is with the weakly interacting meson gas in the low temperature
regime and another is with perturbative Quark Gluon Plasma (QGP) in the high temperature regime. Lattice studies of QCD
Equation of State (EOS) studies~\cite{lap583} indicate that the Steven-Boltzman ideal gas limit can be reached only at
high $T$, and deconfined partons might be strongly interacting in the intermediate temperature range above $T_c$. This
point is supported both by RHIC experiments and theoretically studies. On the one hand, the QGP observed by RHIC can be
well described by the hydrodynamical model\cite{prl86}. On the other hand, the lattice studies of meson correlators
show that charmonia and light hadrons can survive in a temperature range beyond $T_c$~\cite{prl92,prd69,npa715}.

Since glueballs are predicted by QCD and well defined in pure Yang-Mills theory, their $T$-evolution is also a good
probe to investigate the property of QCD matter in the deconfinement phase. In this work, we carry out a numerical
lattice study on anisotropic lattices with much finer lattice in the temporal direction than in spatial ones. By
varying the temporal extension of the lattice, we obtain a wide temperature range from $0.3T_c$ to $1.9T_c$. At each
temperature, we take into account all the twenty $R^{PC}$ channels, with $PC=++,+-,-+,--$ the various parity-charge
conjugate and $R=A_1, A_2, E, T_1, T_2$ the irreducible representations of lattice symmetry group. For each $R^{PC}$,
as in the zero temperature case~\cite{prd56,prd60,prd73}, we implement smearing schemes and the variational method to
acquire an optimal glueball operator which couples most to the ground state. In the data analysis stage, the
correlators of these optimized operators are analyzed through two approaches. First, the thermal masses $M_G$ of
glueballs are extracted in all the channels and all over the temperature range by fitting the correlators with a
single-cosh function form, as is done in the standard hadron mass measurements. Thus the $T$-evolution of the thermal
glueball spectrums are obtained. Secondly, with the respect that the finite temperature effects may result in mass
shifts and thermal widths of glueballs, we also analyze the correlators in $A_1^{++}$, $A_1^{-+}$, $E^{++}$, and
$T_2^{++}$ channels with the Breit-Wigner ansatz which assumes these glueballs thermal widths, say, changes $M_G$ into
$\omega_0-i\Gamma$ in the spectral function (see below). It is expected that the temperature dependence of $\omega_0$
and $\Gamma$ can shed some light on the scenario of the QCD transition.

\section{Numerical details}
%\subsection{Lattices and the gauge action}
We adopt the tadpole improved Symanzik¡¯s action, which has been extensively used in the study of glueballs,
\begin{eqnarray}
&& S_{IA}={\beta}{\{\frac{5}{3}\frac{\Omega_{sp}}{{\xi}u_{s}^{4}}
+\frac{4}{3}\frac{\xi\Omega_{tp}}{u_{t}^{2}u_{s}^{2}}
 -\frac{1}{12}\frac{\Omega_{sr}}{{\xi}u_{s}^{6}}
 -\frac{1}{12}\frac{{\xi}\Omega_{str}}{u_{s}^{4}u_{t}^{2}}\}},
\end{eqnarray}
where $\beta$ is related to the bare QCD coupling constant,
$\xi=a_{s}/a_{t}$ is the aspect ratio for anisotropy (we take $\xi
=5$ in this work), $u_{s}$ and $u_{t}$ are the tadpole improvement
parameters of spatial and temporal gauge links, respectively.
Lattices are $24^3\times N_t$ with $\beta=3.2$, and $a_s =
0.0878$fm, and the spatial volume $V\sim (2.1 {\rm fm})^3$.

%\subsection{The deconfinement critical point}
The deconfinement critical point is roughly determined by the susceptibility $\chi_P$ of Polyakov loops,
\begin{equation}
\chi_P =\langle\Theta ^{2}\rangle-\langle \Theta \rangle^{2}
\end{equation}
where $\Theta$ denotes the $Z(3)$ rotated Polyakov line,
\begin{eqnarray}
\Theta &=&\left\{ \
\begin{array}{ll}
{\rm Re}P\exp[-2\pi i/3];& \arg P\in \lbrack  \pi/3,\pi  ) \\
{\rm Re}P ;              & \arg P\in \lbrack -\pi/3,\pi/3) \\
{\rm Re}P\exp[ 2\pi i/3];& \arg P\in \lbrack -\pi, -\pi/3)
\end{array}
\right..
\end{eqnarray}

\begin{figure}[th]
\centering
\includegraphics[height=5cm]{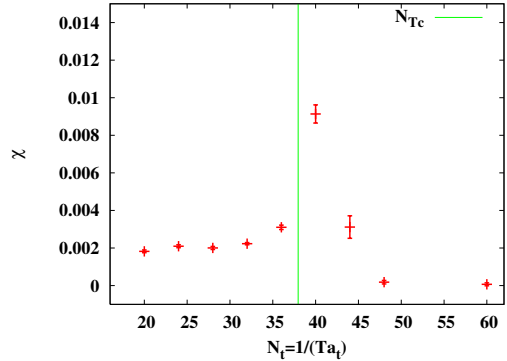}
\caption{$\chi_P$ is plotted versus $N_t$ at $\beta=3.2$. There is a
peak of $\chi_P$ near $N_t=40$.  \label{fig_chi}}
\end{figure}

As illustrated in Fig.~\ref{fig_chi}, the peak position is roughly at $N_t=38$, which corresponds to the critical
temperature $T_c$. With the lattice spacing $a_s=0.0878$fm, $T_c$ is estimated to be $T_c=296$MeV. Based on this, we
vary $N_t$ to obtain a temperature range from $T\sim 0.3 T_c$ to $1.9T_c$, as listed in Table~\ref{simu_para}.

%\subsection{Running parameters}

For all the 20 $J^{PC}$ channels of glueballs, we take the following two steps to construct the optimal glueball
operators which couple most to the ground states (More details can be found in Ref.~\cite{clqcd} and
Ref.~\cite{prd60,prd73}). First, for a given gauge configuration, we generate six differently smeared copies, on each
of which, four realizations of each $J^{PC}$ are established based on all the different spatially oriented Wilson loops
of a set of loop prototypes~\cite{prd60}. As a result, we obtain a set of 24 different glueball operators,
$\phi=\{\phi_\alpha, \alpha=1,2,\ldots,24\}$, for each $J^{PC}$. At the second step, we carry out the variational
method on each operator set $\phi$ to determine the specific combinational coefficients $\{v_\alpha,
\alpha=1,2,\ldots,24\}$ relevant to the ground state, such that the desired optimal operator is obtained as
$\Phi=\sum\limits v_\alpha \phi_\alpha$. Practically at each temperature, after a thermalization of 10000 heatbath
sweeps, the glueball operators are measured every three compound sweeps, each of which is composed of one heatbath and
five micro-canonical over-relaxation(OR) sweeps. In the data analysis, the measurements are divided into $N_{\rm bin}$
bins of the size $n_{\rm mb}=400$ . Parameters $N_{\rm bin}$ and $n_{mb}$ at various temperature are also listed in
Table~\ref{simu_para}.
\begin{table}[h]
\centering

%\begin{ruledtabular}
\begin{tabular}{|c|c|c|c||c|c|c|c|}
\hline
   $N_t$    &   $T/T_{c}$   &   $n_{\rm mb}$    &   $N_{\rm bin}$  &   $N_t$    &   $T/T_{c}$   &   $n_{\rm mb}$    &   $N_{\rm bin}$  \\
  \hline
    128     &   0.30        &   400          &   24  &  36      &   1.05        &   400          &   40         \\
    80      &   0.47        &   400          &   30  &  32      &   1.19        &   400          &   56         \\
    60      &   0.63        &   400          &   44  &  28      &   1.36        &   400          &   40         \\
    48      &   0.79        &   400          &   40  &  24      &   1.58        &   400          &   40         \\
    44      &   0.86        &   400          &   44  &  20      &   1.90        &   400          &   40         \\
    40      &   0.95        &   400          &   40  &  --      &   --          &   --           &   --         \\
   \hline
\end{tabular}
\caption{\label{simu_para}Temperature range and simulation parameters in this work.}
%\end{ruledtabular}
\end{table}

\section{Glueballs at finite temperature}
Theoretically, under the periodic boundary condition in the temporal direction, the temporal correlators $C(t,T)$ at
the temperature $T$ can be written in the spectral representation as
\begin{eqnarray}
\label{aaa} C(t,T)&\equiv& \frac{1}{Z(T)}{\rm Tr} \left( e^{-
H/T}\Phi(t)\Phi(0)\right)\nonumber \\
&=&\sum\limits_{m,n}\frac{|\langle n|\Phi|m\rangle|^2} {2Z(T)} \exp \left(-\frac{E_m+E_n}{2T}\right)\nonumber \times
\cosh\left[\left(t-\frac{1}{2T}\right)(E_n-E_m)\right]\nonumber\\
&=&\int\limits_{-\infty}^{\infty} d\omega \rho(\omega) K(\omega,T),
\end{eqnarray}
with a $T$-dependent kernel
%\begin{equation}
$K(\omega,T)=\frac{\cosh(\omega/(2T)-\omega t)}{\sinh(\omega/(2T))}$
%\end{equation}
and the spectral function,
\begin{equation}
\label{spectral}
 \rho(\omega)=\sum\limits_{m,n}\frac{|\langle
n|\Phi|m\rangle|^2}{2Z(T)}e^{-E_m/T}(\delta(\omega-(E_n-E_m)-\delta(\omega-(E_m-E_n)),
\end{equation}
where $Z(T)$ is the partition function at $T$, and $E_n$ the energy of the thermal state $|n\rangle$ ($|0\rangle$
represents the vacuum state). In the zero-temperature limit($T\rightarrow 0$), due to the factor $\exp(-E_m/T)$, the
spectral function $\rho(\omega)$ degenerates to
\begin{equation}
\label{single_pole} \rho(\omega)=\sum\limits_n\frac{|\langle 0|\Phi|n\rangle|^2}{2Z(0)}
\left(\delta(\omega-E_n)-\delta(\omega+E_n)\right),
\end{equation}
thus we have the function form of the correlation function,
%\begin{equation}
$C(t,T=0)=\sum\limits_n W_n e^{-E_n t}$
%\end{equation}
with $W_n=|\langle 0|\Phi|n\rangle|^2/Z(0)$. However, for any finite temperature (this is always the case for finite
lattices), all the thermal states with the non-zero matrix elements $\langle m|\Phi|n\rangle$ may contribute to the
spectral function $\rho(\omega)$. Intuitively in the confinement phase, the fundamental degrees of freedom are
hadron-like modes, thus the thermal states should be multi-hadron states. If they interact weakly with each other, we
can treat them as free particles at the lowest order approximation and consider $E_m$ as the sum of the energies of
hadrons including in the thermal state $|m\rangle$. Since the contribution of a thermal state $|m\rangle$ to the
spectral function is weighted by the factor $\exp(-E_m/T)$, apart from the vacuum state, the maximal value of this
factor is $\exp(-M_{\rm min}/T)$ with $M_{\rm min}$ the mass of the lightest hadron mode in the system. As far as the
quenched glueball system is concerned, the lightest glueball is the scalar, whose mass at the low temperature is
roughly $M_{0^{++}}\sim 1.6$ GeV, which gives a very tiny weight factor $\exp(-M_{0^{++}}/T_c)\sim 0.003$ at $T_c$ in
comparison with unity factor of the vacuum state. That is to say, for the quenched glueballs, up to the critical
temperature $T_c$, the contribution of higher spectral components beyond the vacuum to the spectral function are much
smaller than the statistical errors (the relative statistical errors of the thermal glueball correlators are always a
few percents) and can be neglected. As a result, the function form of $\rho(\omega)$ in Eq.~\ref{single_pole} can be a
good approximation for the spectral function of glueballs at least up to $T_c$. Accordingly, considering the finite
extension of the lattice in the temporal direction, the function form of the thermal correlators can be approximated as
\begin{equation}
\label{cosh_function} C(t,T)=\sum\limits_n W_n \frac{\cosh(M_n(1/(2T)-t))}{\sinh(M_n/(2T))},
\end{equation}
which is surely the commonly used function form for the study of hadron masses at low temperatures on the lattice. As
is always done, the glueball masses $M_n$ derived by this function are called the pole masses in this work.

\subsection{The single-cosh fit}

\begin{figure}[thb]
  \begin{center}
    \begin{tabular}{ccc}
      \resizebox{45mm}{!}{\includegraphics{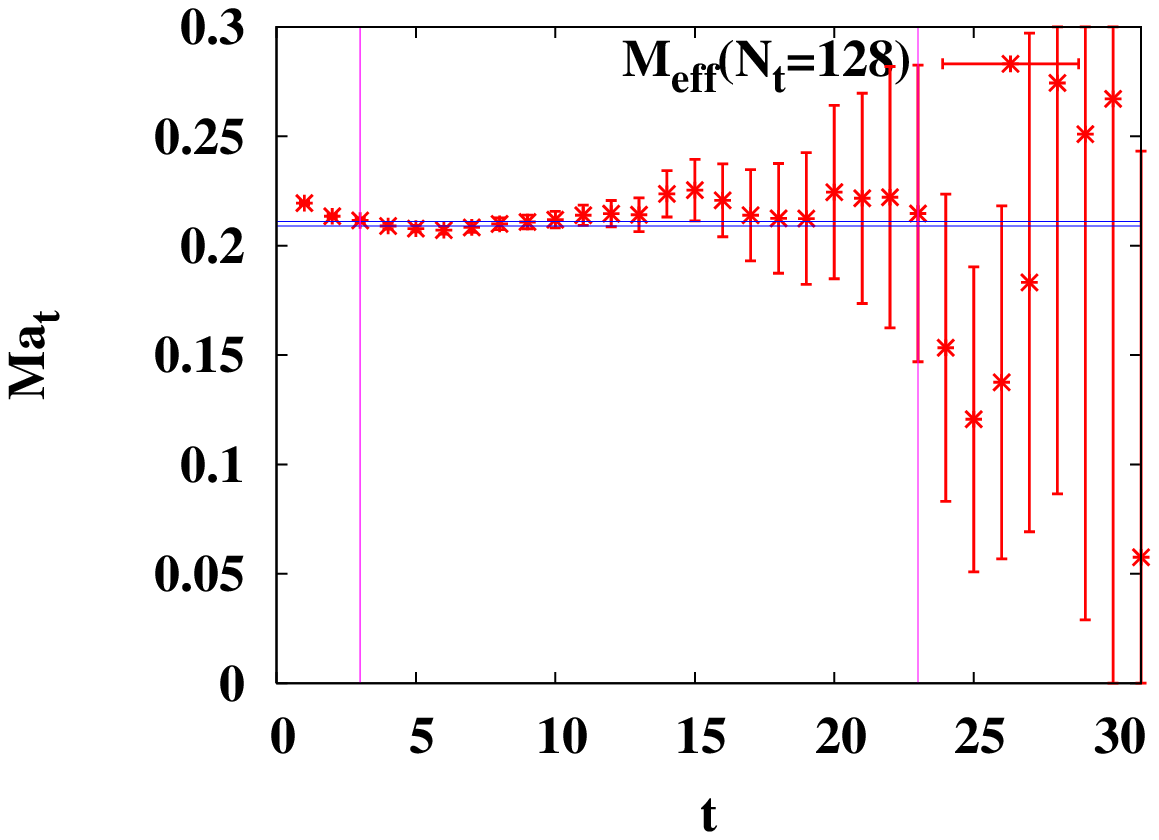}} &
      \resizebox{45mm}{!}{\includegraphics{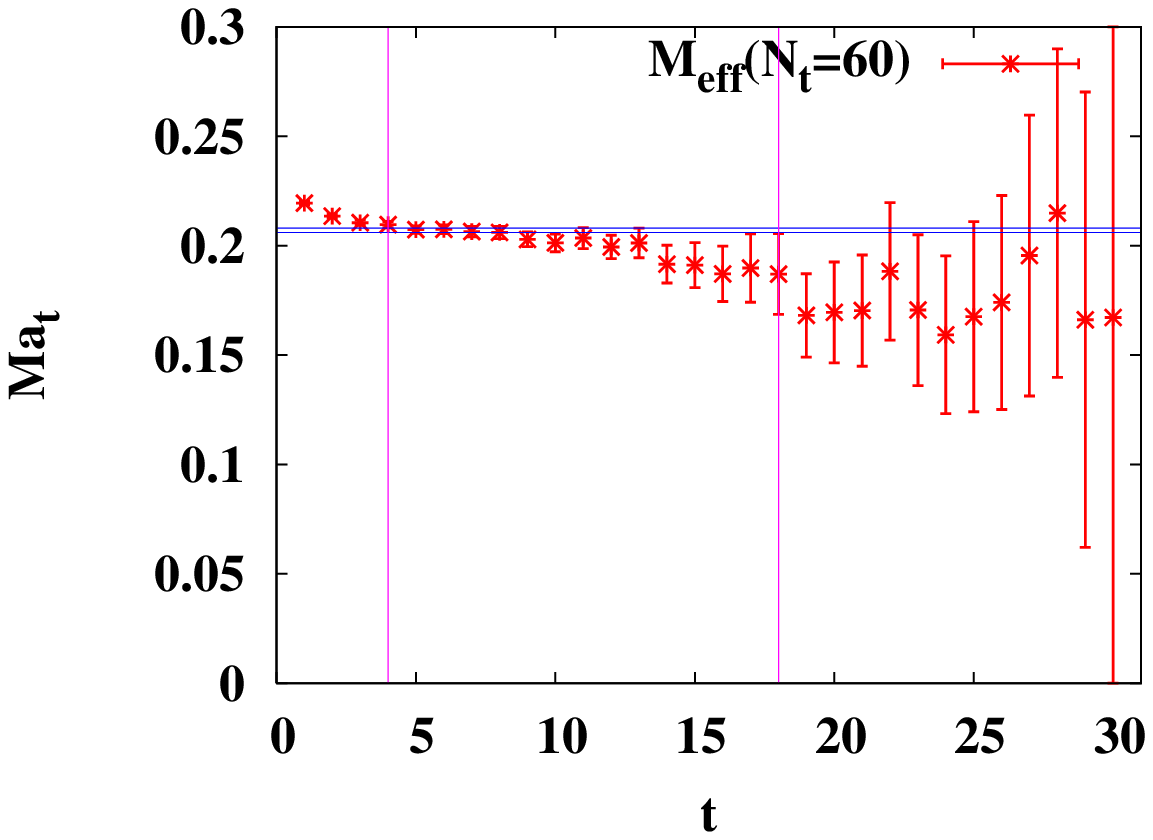}} &
      \resizebox{45mm}{!}{\includegraphics{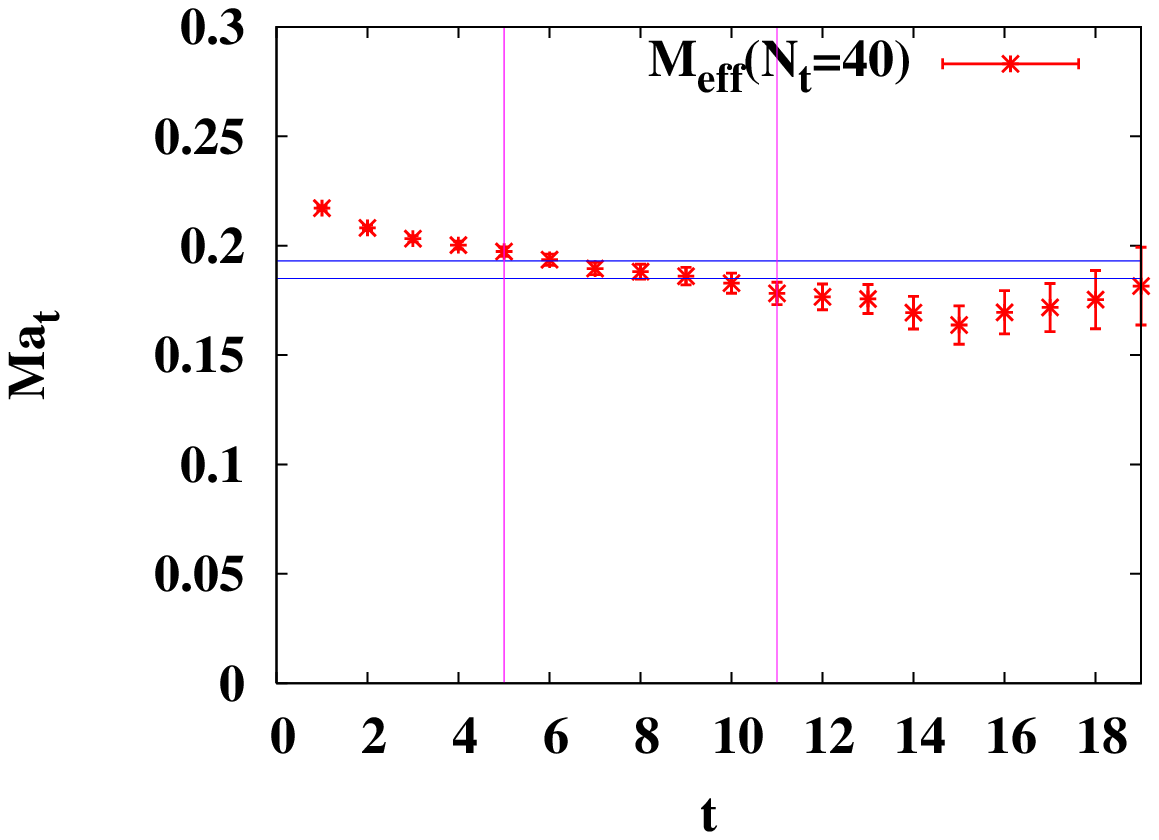}} \\
      \resizebox{45mm}{!}{\includegraphics{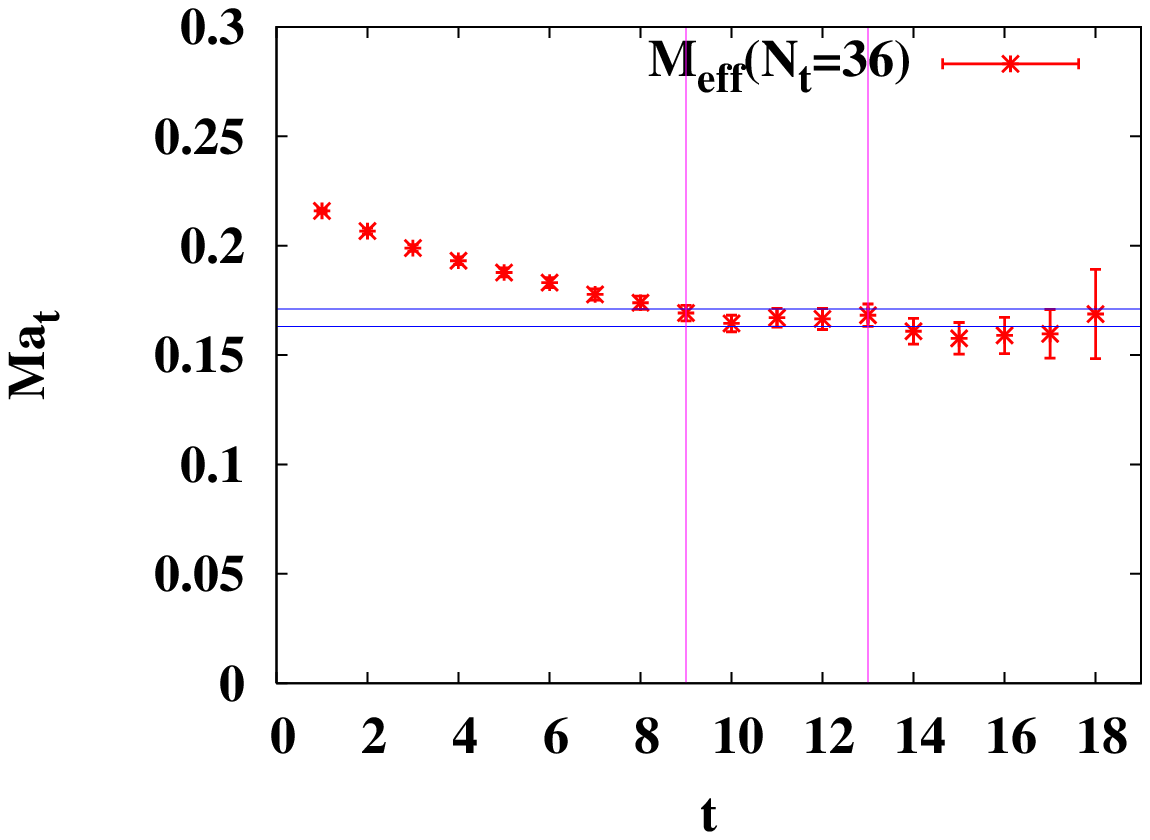}}&
      \resizebox{45mm}{!}{\includegraphics{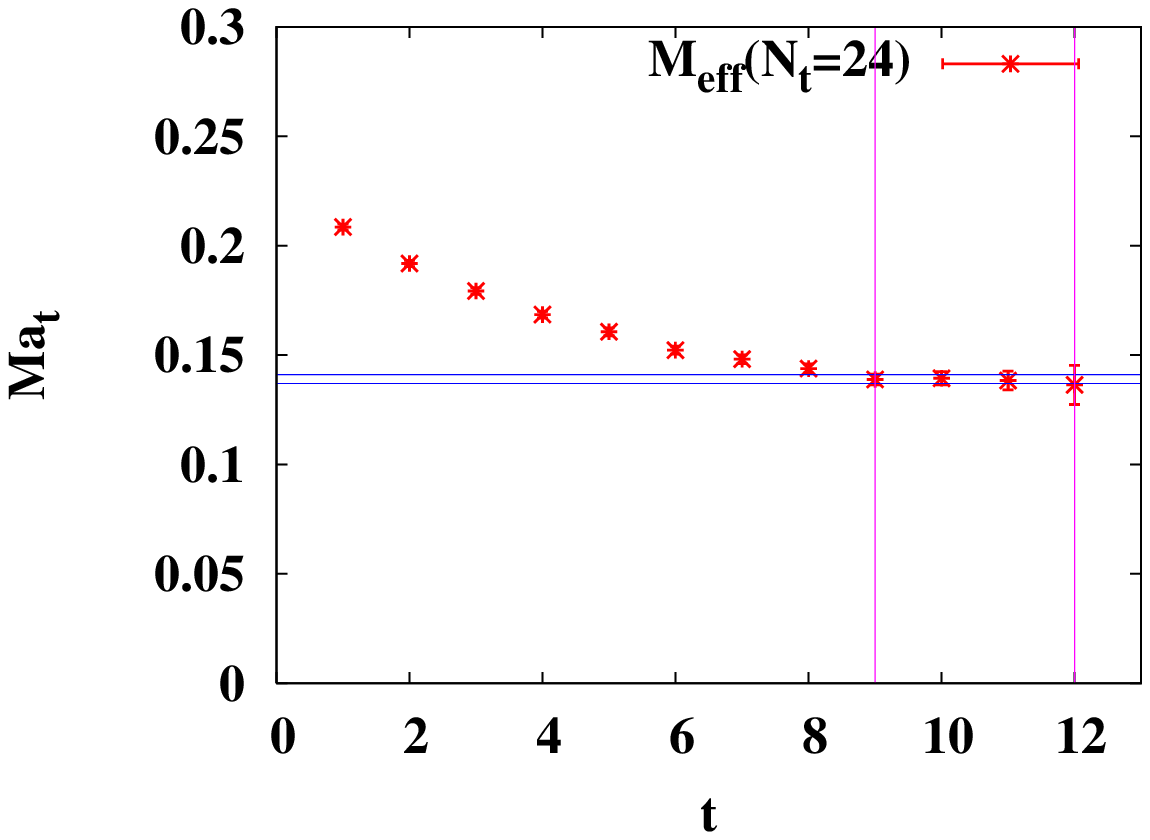}} &
      \resizebox{45mm}{!}{\includegraphics{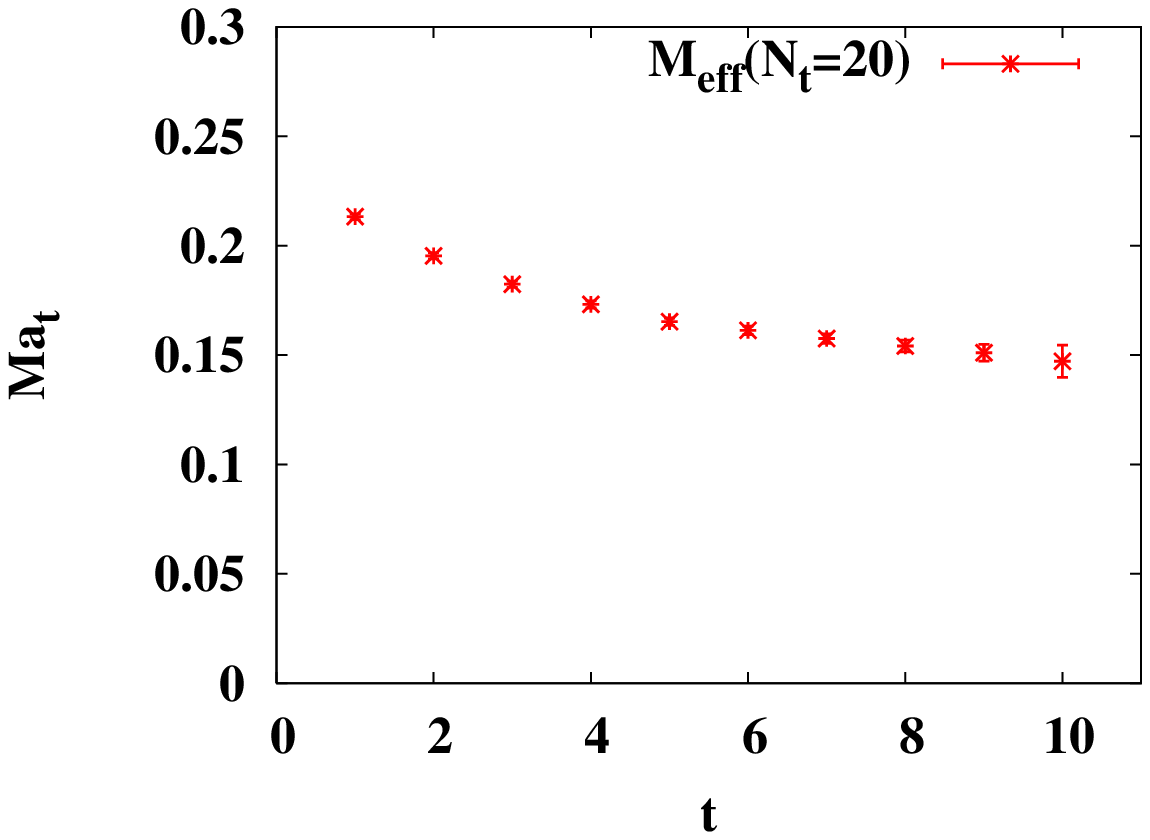}}
    \end{tabular}
    \caption{Effective masses at different temperatures in $E^{++}$ channel.
    Data points are the effective masses with jackknife error bars.
    The vertical lines indicate the time window $[t_1, t_2]$ over which the
    single-cosh fittings are carried out, while the horizonal lines illustrate
    the best fit result of pole masses (in each panel the double horizonal lines
    represent the error band estimated by jackknife analysis)}
    \label{opt09}
  \end{center}
\end{figure}

After the thermal correlators $C(t, T)$ are obtained, the pole masses of the ground state (or the lowest spectral
component) can be extracted straightforwardly. For each $R^{PC}$ channel and at each temperature $T$, we first
calculate the effective mass $M_{\rm eff}(t)$ as a function of $t$ by solving the equation
\begin{equation}
\frac{C(t+1,T)}{C(t,T)}=\frac{\cosh ( (t + 1 - N_{t}/2)a_{t}M_{\rm eff}(t) )}{\cosh ( (t - N_{t}/2)a_{t}M_{\rm eff}(t)
)}, \label{cosh},
\end{equation}
and determine the time window $[t_1,t_2]$ where $M_{\rm eff}$(t) has a plateau. In this time window, $C(t,T)$ is fitted
through a single-cosh function form. Fig.~\ref{opt09} illustrates this procedure in $E^{++}$ channel.

\begin{figure}
\centering
\includegraphics[height=5cm]{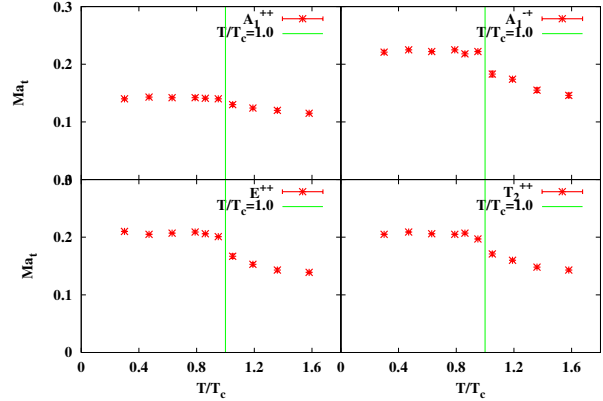}
\caption{The $T$-dependence of pole masses $A_{1}^{++}$,
$A_{1}^{-+}$ $E^{++}$, and $T_{2}^{++}$ glueballs.
\label{opt_total}}
\end{figure}

In this work, the pole masses in all the 20 $R^{PC}$ channels are extracted at all the temperatures. The common feature
of temperature dependence of pole masses is that, below $T_c$, the pole masses keep stable when varying the
temperature, while above $T_c$, the pole masses decrease gradually and cannot be extracted beyond the temperature
$T=1.6T_c$. Fig.~\ref{opt_total} illustrates these hebaviors in $A_1^{++}$, $A_1^{-+}$, $E^{++}$, and $T_2^{++}$
channels. These results are different from the observation of a previous lattice study on glueballs where the observed
pole-mass reduction start even at $T\simeq 0.8 T_{c}$~\cite{prd66}.

\subsection{Breit-Wigner analysis (BW)}
%The result of the sing-cosh analysis is in agreement with the picture that the state of matter below $T_c$ are made up
%of weakly interacting glueball-like modes. While when $T>T_c$, the thermal correlators deviate from
%Eq.~\ref{cosh_function} more and more. This observation implies the degrees of freedom are very different from that
%when $T<T_c$.
Theoretically in the deconfined phase, gluons can be liberated from hadrons. However, the study of the
equation of state shows that the state of the matter right above $T_c$ is far from a perturbative gluon gas. In other
words, the gluons in the intermediate temperature above $T_c$ may interact strongly with each other and glueball-like
resonances can be possibly formed. Thus different from bound states at low temperature, thermal glueballs can acquire
thermal width due to the thermal scattering between strongly interacting gluons and the magnitudes of the thermal
widths can signal the strength of these type of interaction at different temperature.

\begin{figure}[tb]
  \centering
    \begin{tabular}{ccc}
     (a)~$N_t=128$($T/T_c=0.32$) & (b)~$N_t=36$($T/T_c=1.09$) &(c)~$N_t=20$($T/T_c=1.97$)\\
      \resizebox{40mm}{!}{\includegraphics{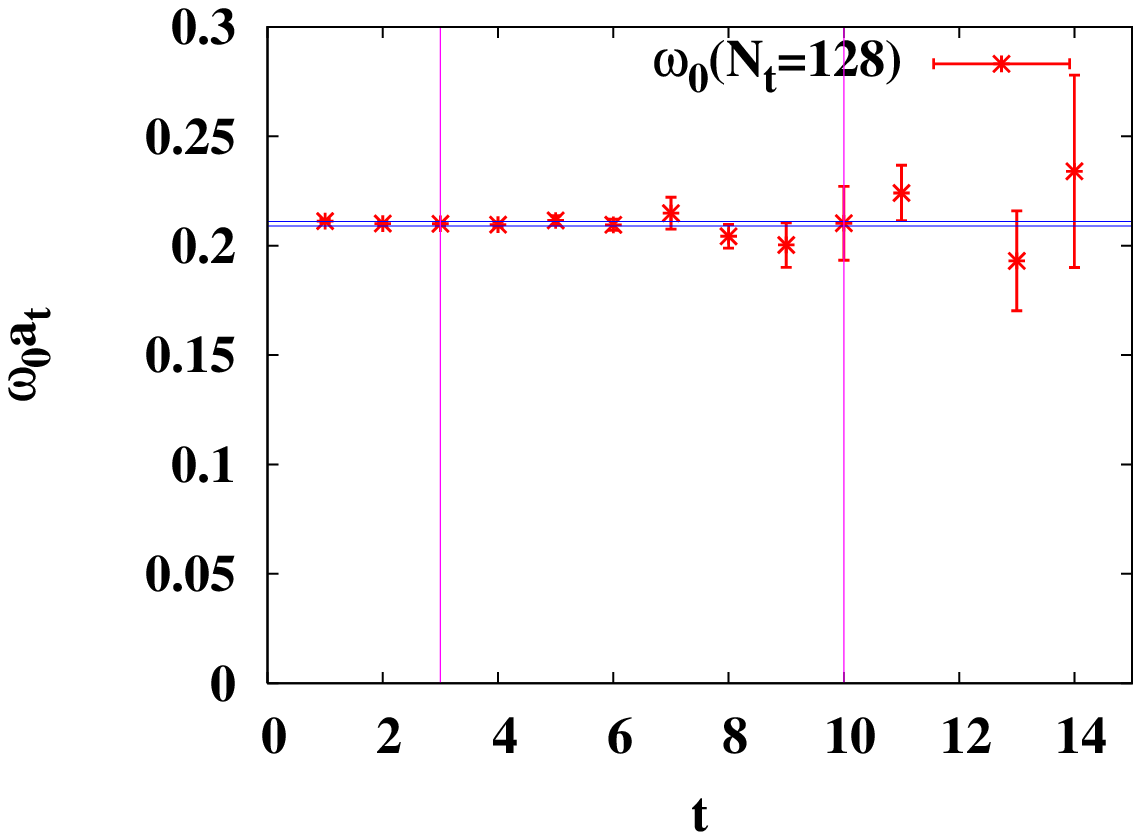}} &
      \resizebox{40mm}{!}{\includegraphics{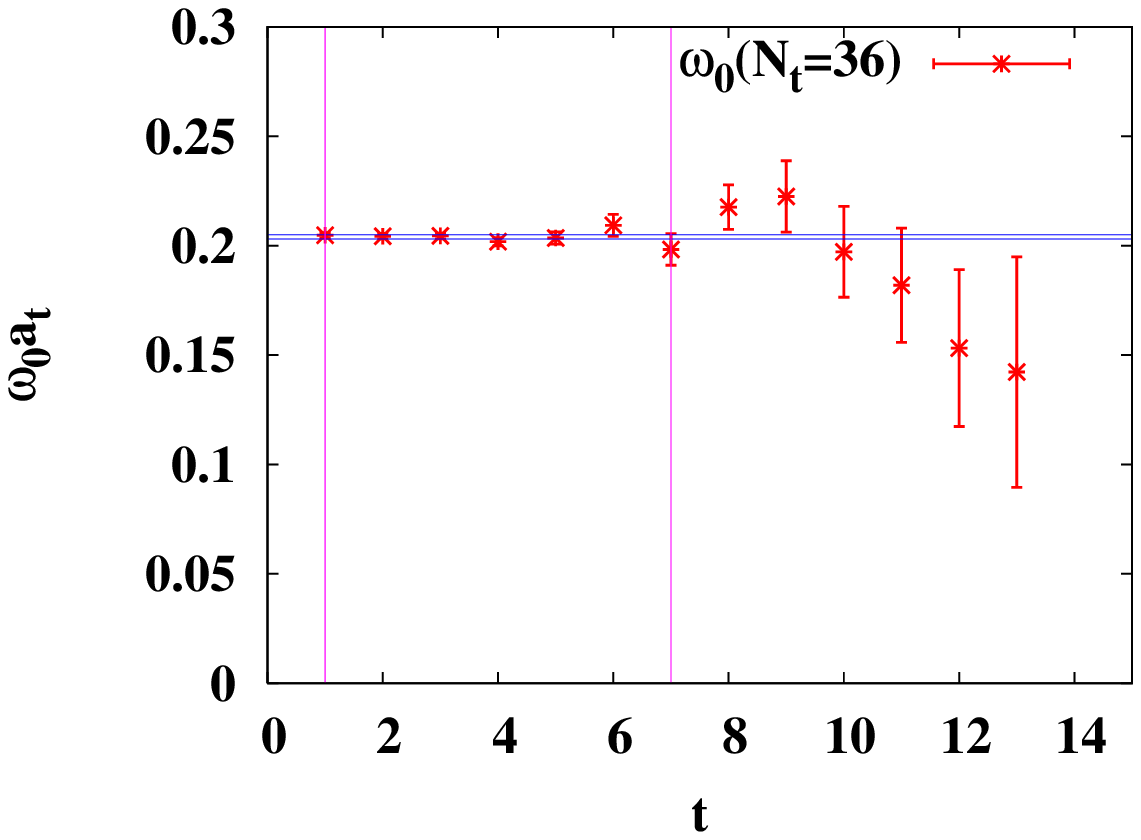}} &
      \resizebox{40mm}{!}{\includegraphics{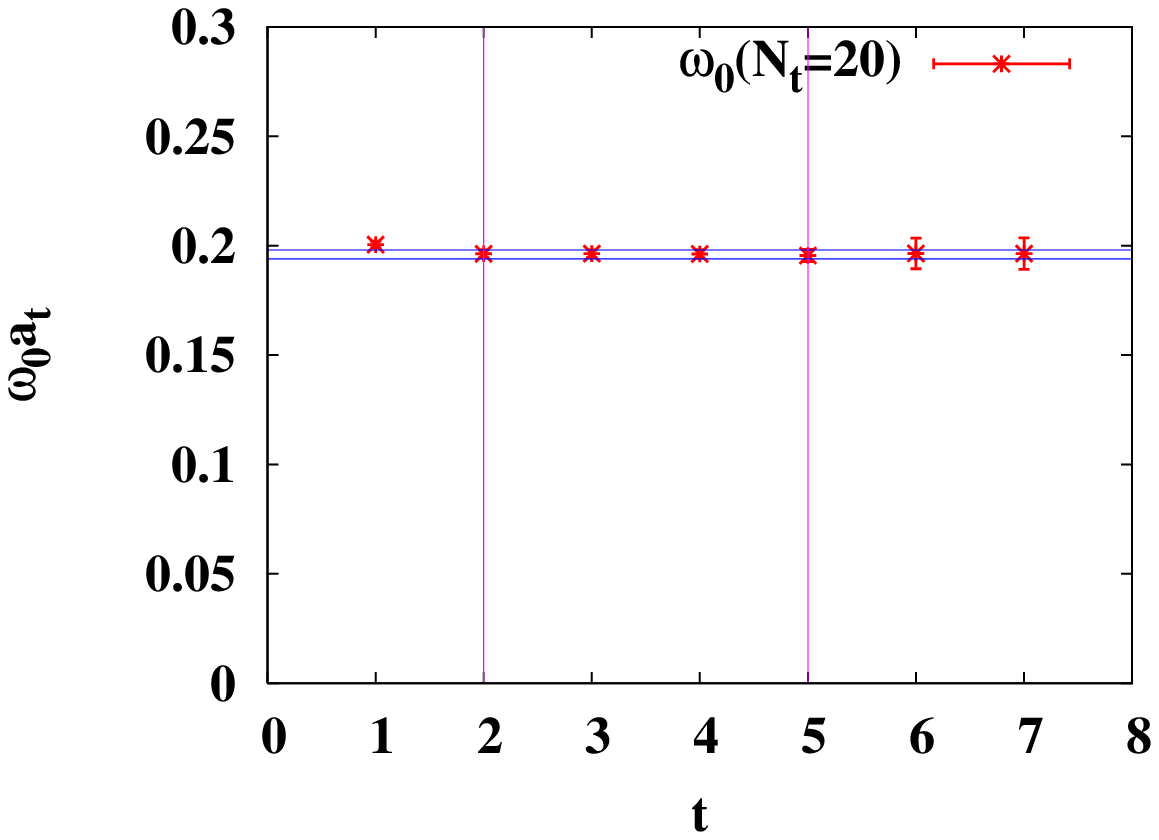}} \\
      \resizebox{40mm}{!}{\includegraphics{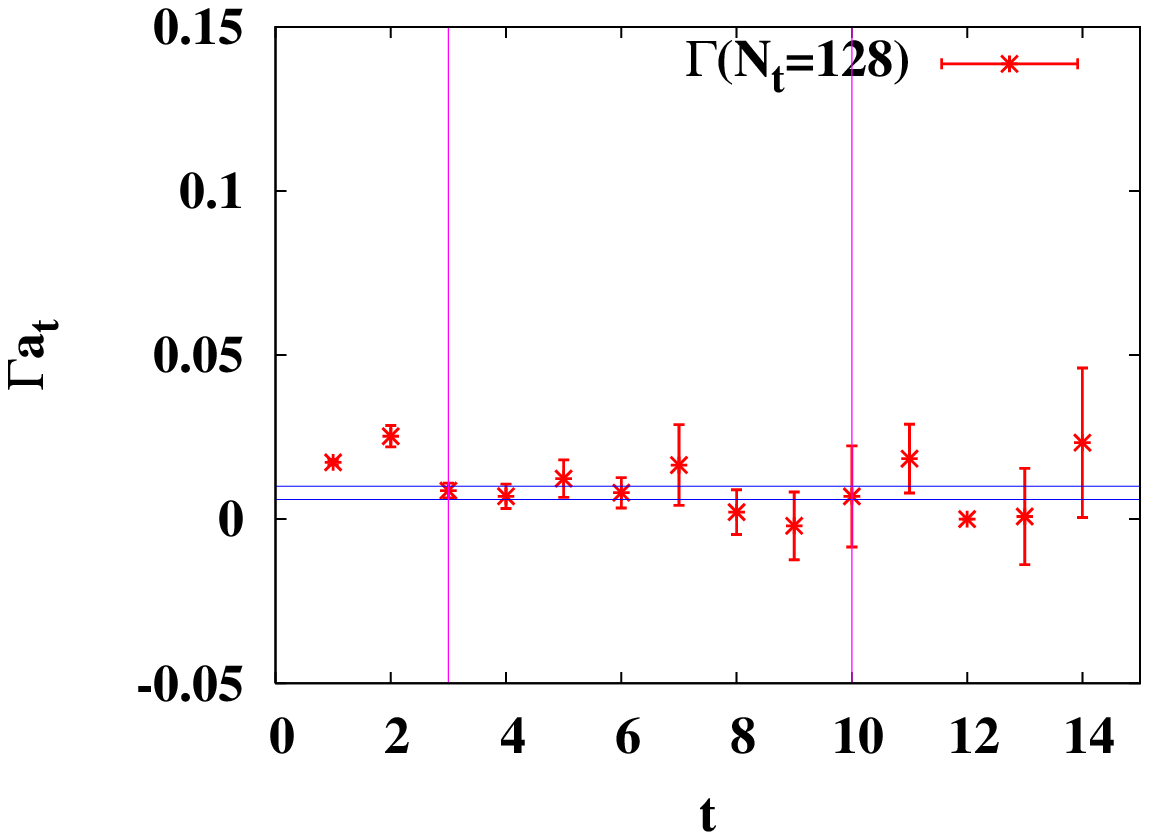}} &
      \resizebox{40mm}{!}{\includegraphics{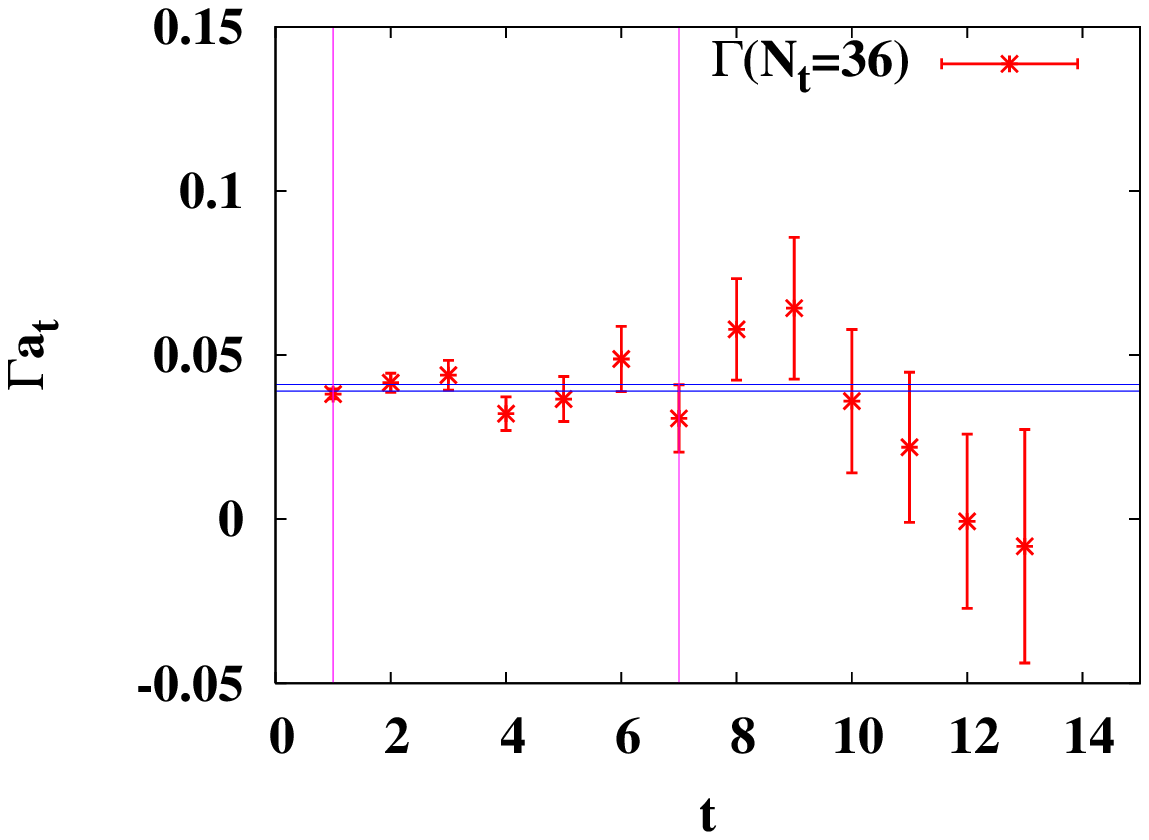}}  &
      \resizebox{40mm}{!}{\includegraphics{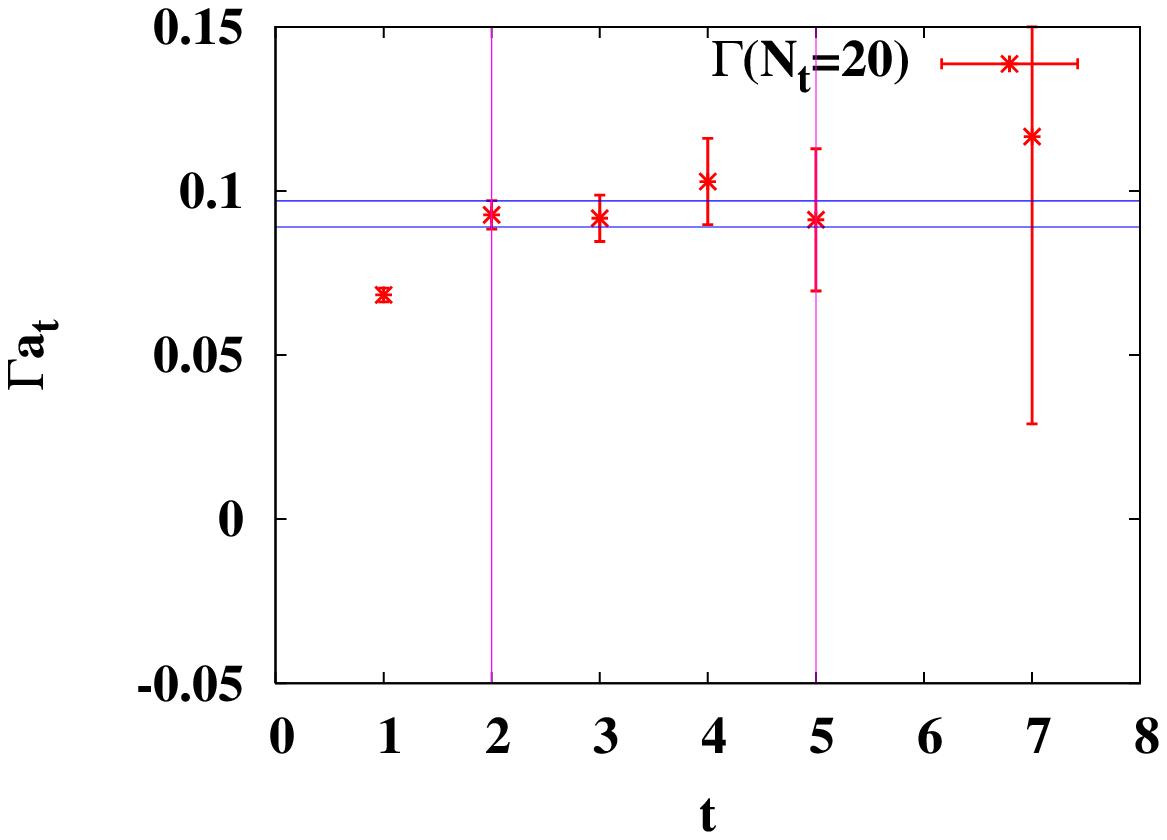}}
    \end{tabular}
    \caption{ Determinations of the fit range $[t_1,t_2]$ in $T_2^{++}$ channel at $N_t=$ 128, 36, and 20. In each
          row, $\omega_0^{\rm (eff)}(t)$ and $\Gamma^{\rm (eff)}(t)$ are plotted by data points with jackknife error bars.
          $[t_1,t_2]$ are chosen to include the time slices
          between the two vertical lines, where $\omega_0^{\rm (eff)}(t)$ and $\Gamma^{\rm (eff)}(t)$ show up plateaus
          simultaneously. The best fit results of $\omega_0$ and $\Gamma$ through the function $g_\Gamma(t)$
          are illustrated by the horizonal lines.\label{opt17_bw}
          }
\end{figure}

By assuming glueballs thermal widths, we also adopt the Breit-Wigner ansatz~\cite{prd66} to analyze the thermal
correlators once more. First, we treat thermal glueballs as resonance objects which correspond to the poles (denoted by
$\omega=\omega_0-i\Gamma$) of the retarded and advanced Green functions in the complex $\omega-$plane. $\omega_0$ is
called the mass of the resonance glueball and $\Gamma$ its thermal width in this work.  Secondly, we assume that the
spectral function $\rho(\omega)$ is dominated by these resonance glueballs. With these respects, the thermal
correlators can be parameterized as~\cite{clqcd,prd66}
\begin{eqnarray}
%\begin{equation}
\label{fit_fun} g_\Gamma(t)&=&A\left[ {\rm Re} \left( \frac{\cosh((\omega_0+i\Gamma)(\frac{1}{2T}-t))}
{\sinh(\frac{(\omega_0+i\Gamma)}{2T})} \right) \right. \nonumber \\
 &+&
 \left. 2\omega_0 T\sum\limits_{n=1}^{\infty}\cos\left(2\pi nTt \right)
\left\{ \frac{1}{(2\pi nT+\Gamma)^2+\omega_0^2}-(n\rightarrow -n)
 \right\}
 \right].
%\end{equation}
\end{eqnarray}
In the data analysis, the effective mass $\omega_0^{\rm{eff}}(t)$ and effective width $\Gamma^{\rm{eff}}(t)$ are
obtained first by solving the equation array,
\begin{eqnarray}
\label{solution} {g_\Gamma(t)}/{g_\Gamma(t+1)}&=& {C(t,T)}/{C(t+1,T)}\nonumber\\
{g_\Gamma(t+1)}/{g_\Gamma(t+2)}&=&{C(t+1,T)}/{C(t+2,T)},
\end{eqnarray}
where $C(t,T)$ is the measured correlator, then the simultaneous plateau region of $\omega_0(t)$ and $\Gamma(t)$ gives
the fit window $[t_1,t_2]$ where the fit is carried out. This procedure in $T_2^{++}$ channel is shown in
Fig.~\ref{opt17_bw} for instance.

\begin{figure*}[ht]
\centering
\begin{tabular}{lcr}
     \resizebox{55mm}{!}{\includegraphics{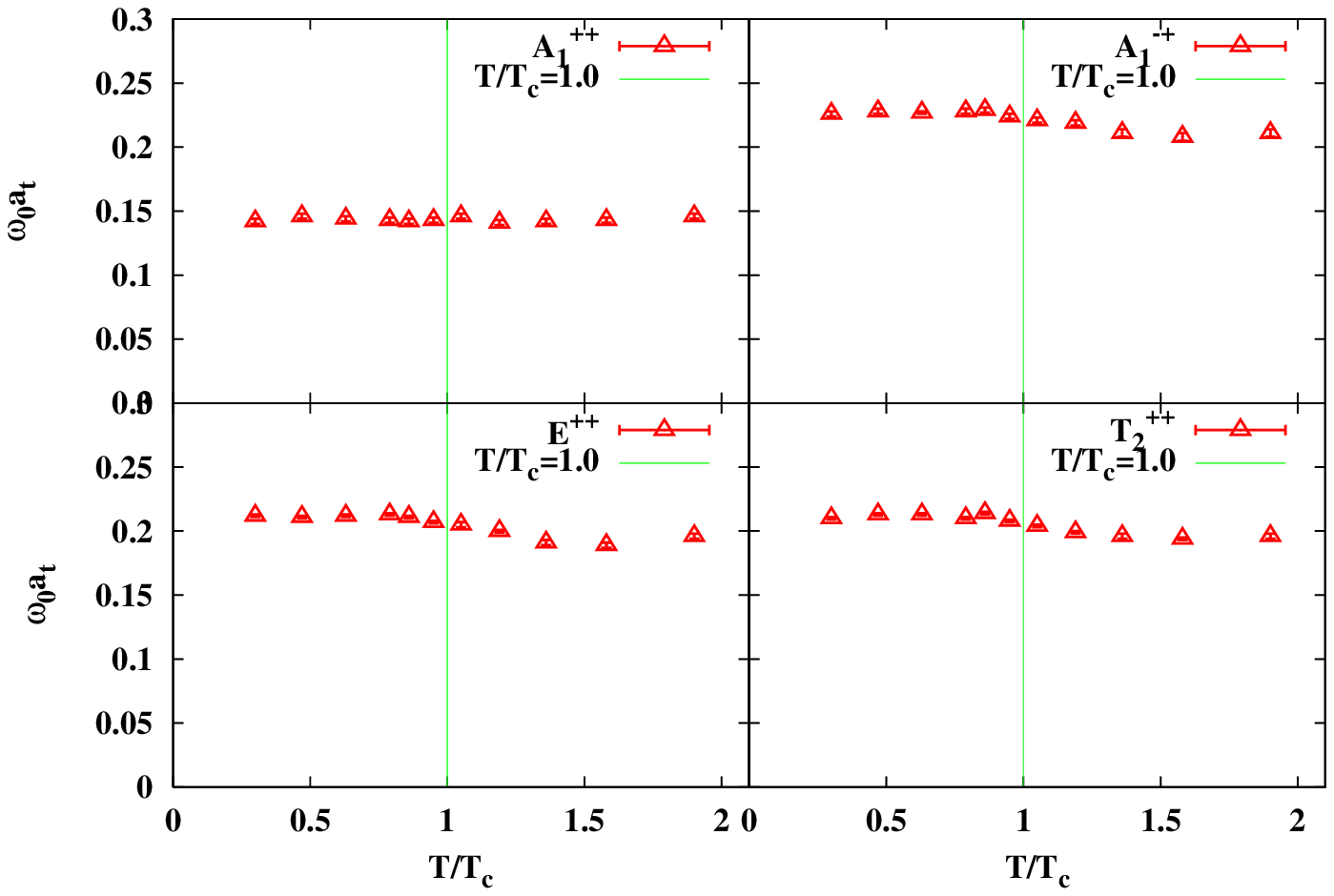}} &&
     \resizebox{55mm}{!}{\includegraphics{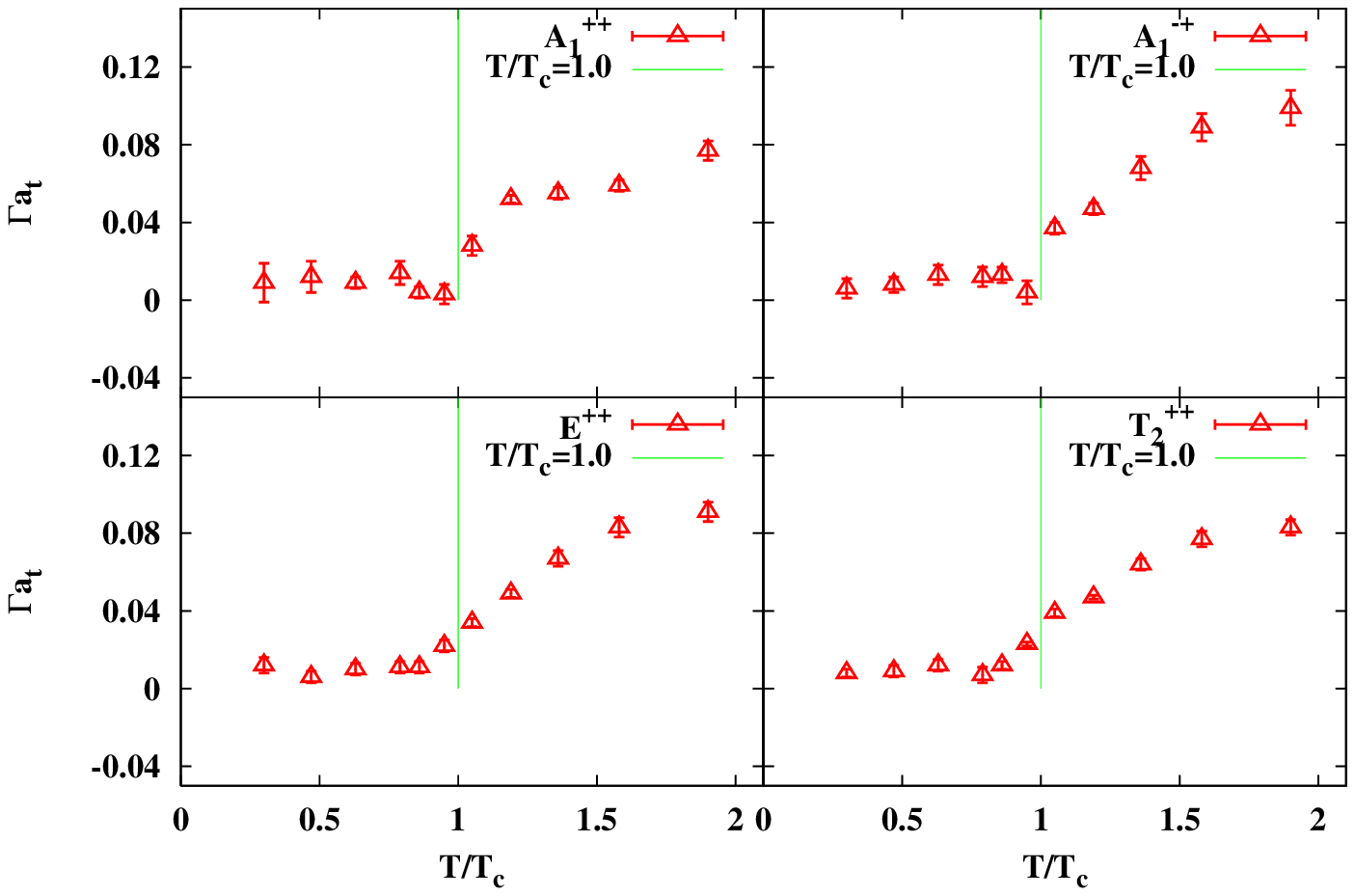}}

\end{tabular}
\caption{$\omega_0$'s and $\Gamma$'s are plotted versus $T/T_c$ for
$A_{1}^{++}$, $A_{1}^{-+}$ $E^{++}$, and $T_{2}^{++}$ channels. The
vertical lines indicate the critical temperature.
\label{opt_total_om}}
\end{figure*}

The main feature of the best fit $\omega_0$ and $\Gamma$ in $A_1^{++},A_1^{-+},E^{++}$, and $T_2^{++}$ channels is
illustrated in Figure~\ref{opt_total_om}: First, $\omega_0$'s are insensitive to $T$, or more specifically, the
reduction of $\omega_0$ at the highest temperature $T=1.90T_c$ are less than $5\%$. Secondly, $\Gamma$'s are small and
do not vary much below $T_c$, but increase abruptly when the temperature passes $T_c$ and reach to values $\sim
\omega_0/2$ at $T=1.90T_c$. These features can be easily seen in Fig.~\ref{opt_total_om}, where the behaviors of
$\omega_0$ and $\Gamma$ with respect to the temperature $T$ are plotted for all the four channels.

\section{Summary and conclusion}
In the pure SU(3) gauge theory, the thermal correlators in all the 20 symmetry channels are calculated on anisotropic
lattices in the temperature range from 0.3 $T_c$ to 1.9$T_c$. Both the single-cosh fit and BW analysis show that
glueballs can survive up to 1.9$T_c$. Our results are consistent with that of the studies of EOS and charmonia
~\cite{prl92,prd69,soft_modes,quark_number}. In BW analysis, glueball masses keep stable when the temperature
increasing, and the thermal widths of glueballs becomes larger and larger above $T_c$. It seems that in the
intermediate T range, the state of matter are dominated by strongly interacting gluons. Gluons interact with each other
strongly enough to form glueball-like resonances, in the mean time, glueballs can also decay into gluons. At a given
temperature, these two procedure reach the thermal equilibrium. The thermal widths signal the interaction strength.

\vskip 0.5cm This work is supported in part by NSFC (Grant No. 10575107, 10675005, 10675101, 10721063, and 10835002)
and CAS (Grant No. KJCX3-SYW-N2 and KJCX2-YW-N29). The numerical calculations were performed on DeepComp 6800
supercomputer of the Supercomputing Center of Chinese Academy of Sciences, Dawning 4000A supercomputer of Shanghai
Supercomputing Center, and NKstar2 Supercomputer of Nankai University.

%%%%%%%%%%%%%%%%%%%%%%%%%%%%%%%%%%%%%%%%%%%%%%%%%%%%%%%%%%%%%%%%%%%%%%%%%%%%%%%%%%%%%%%%%%%%%%%%%%%%%%%%%%%%%%%%%%%%%%%%%%%%%%%%%%%%%%%%%%%%%%%%%%%%%%%%

%%%%%%%%%%%%%%%%%%%%%%%%%%%%%%%%%%


\begin{thebibliography}{10}
\bibitem{lap583} P. Petreczky, hep-lat/0409139; F. Karsch, Lect. Notes Phys. {\bf 583}, 209 (2002), {\tt
 arXiv:hep-lat/0106019}; D.E. Miller, Acta Physica Polonica B {\bf 28}, 2937 (1997), {\tt arXiv:hep-ph/9807304}.
\bibitem{prl86} STAR Collaboration: K.H. Ackermann, {\it et al.}, Phys. Rev. Lett. {\bf 86}, 402 (2001),
{\tt arXiv:nucl-ex/0009011 }.
\bibitem{prl92} M. Asakawa and T. Hatsuda, Phys. Rev. Lett. {\bf 92}, 012001 (2004), {\tt arXiv:hep-lat/0308034}.
\bibitem{prd69} S. Datta, F. Karsch, P. Petreczky, and I. Wetzorke, Phys. Rev. D {\bf 69}, 094507 (2004),
     {\tt arXiv:hep-lat/0312037}; S. Datta, F. Karsch, P. Petreczky, and I. Wetzorke, J.Phys. G {\bf 30}, S1347 (2004).
\bibitem{npa715}M. Asakawa {\it et al.}, Nucl. Phys. A {\bf 715} 863 (2003).
\bibitem{prd56} C.J. Morningstar and M. Peardon, Phys. Rev. D {\bf 56}, 4043 (1997).
\bibitem{prd60} C.J. Morningstar and M. Peardon, Phys. Rev. D {\bf 60}, 034509 (1999).
\bibitem{prd73} Y. Chen {\it et al.}, Phys. Rev. D {\bf 73}, 014516 (2006).
\bibitem{clqcd} X.-F. Meng {\it et al.} [CLQCD Collaboration], {\tt arXiv:0903.1991 [hep-lat]}
\bibitem{prd66} N. Ishii, H. Suganuma, and H. Matsufuru, Phys. Rev. D {\bf 66}, 094506 (2002), {\tt arXiv:hep-lat/0206020
}; N. Ishii, H. Suganuma, and H. Matsufuru, Phys. Rev. D {\bf 66},
014507 (2002), {\tt arXiv:hep-lat/0309102}.
\bibitem{mpla21} W. Liu {\it et al.}, Mod. Phys. Lett. A {\bf 21}, 2313 (2006).
\bibitem{soft_modes}S. Gottlieb {\it et al.}, Phys. Rev. Lett.
             {\bf 59}, 2247 (1987); S. Gottlieb {\it et al.}, Phys. Rev. D {\bf 38}, 2888 (1988);
             S. Gottlieb {\it et al.}, Phys. Rev. D {\bf 55}, 6852 (1997).
\bibitem{quark_number} C. DeTar, Phys. Rev. D {\bf 32}, 276 (1985); Phys. Rev. D {\bf 37}, 2328 (1988);
G. Boyd, S. Gupta, F. Karsch, and E. Laermann, Z. Phys. C {\bf 64}, 331 (1994); H. Iida, T. Doi, N. Ishii, H. Suganuma,
and K. Tsumura, Phys. Rev. D {\bf 74}, 074502 (2006), {\tt arXiv:hep-lat/0602008 }; A. Jakovac, P. Petreczky, K.
Petrov, and A. Velytsky, PoS(JHW 2005), {\tt arXiv:hep-lat/0603005}; B. Berg and A. Billoire, Nucl. Phys. A {\bf 783},
477 (2007); Ph. de Forcrand {\it et al.}[QCD-TARO Collaboration], Phys. Rev. D {\bf 63}, 054501 (2001); T. Umeda, R.
Katayama,O. Miyamura and H. Matsufuru, Int. J. Mod. Phys. A {\bf 16}, 2215 (2001).
\end{thebibliography}
\end{document}